\pdfoutput=1

\documentclass[11pt]{article}

\usepackage[final]{acl}

\usepackage{times}
\usepackage{latexsym}
\usepackage{multirow}
\usepackage{booktabs}
\usepackage{amsmath}
\usepackage{amssymb}
\usepackage{cleveref}
\usepackage{graphicx}
\usepackage{epstopdf}
\usepackage{hyperref}

\usepackage[T1]{fontenc}

\usepackage[utf8]{inputenc}

\usepackage{microtype}

\usepackage{inconsolata}

\title{VISTA: Visualized Text Embedding For Universal Multi-Modal Retrieval} 

\author{Junjie Zhou$^1$\thanks{Work done during Junjie's internship at BAAI.},  Zheng Liu$^{2,3}$\thanks{Corresponding author.}, Shitao Xiao$^2$, Bo Zhao$^2$, Yongping Xiong$^1$  \\
        $^1$ State Key Laboratory of Networking and Switching Technology, \\
        Beijing University of Posts and Telecommunications \\ 
        $^2$ Beijing Academy of Artificial Intelligence \\
        $^3$ The Hong Kong Polytechnic University \\
        \texttt{zhoujunjie@bupt.edu.cn} \quad
        \texttt{zhengliu1026@gmail.com} \quad
        \texttt{stxiao@baai.ac.cn} \\
}

\begin{document}
\maketitle
\begin{abstract}
Multi-modal retrieval becomes increasingly popular in practice. However, the existing retrievers are mostly text-oriented, which lack the capability to process visual information. Despite the presence of vision-language models like CLIP, the current methods are severely limited in representing the text-only and image-only data. In this work, we present a new embedding model \textbf{VISTA} for universal multi-modal retrieval. Our work brings forth threefold technical contributions. Firstly, we introduce a flexible architecture which extends a powerful text encoder with the image understanding capability by introducing visual token embeddings. Secondly, we develop two data generation strategies, which bring high-quality composed image-text to facilitate the training of the embedding model. Thirdly, we introduce a multi-stage training algorithm, which first aligns the visual token embedding with the text encoder using massive weakly labeled data, and then develops multi-modal representation capability using the generated composed image-text data. In our experiments, VISTA achieves superior performances across a variety of multi-modal retrieval tasks in both zero-shot and supervised settings. Our model, data, and source code are available at \href{https://github.com/FlagOpen/FlagEmbedding/tree/master/FlagEmbedding/visual}{https://github.com/FlagOpen/FlagEmbedding}. 
\end{abstract} 

\section{Introduction}
Information retrieval (IR) is a critical task in many real-world scenarios, e.g., search engines, open-domain question answering, and retrieval augmented generation \cite{dpr,lewis2020retrieval}. It aims to find relevant data from a large database such that the downstream problems can be faithfully solved on top of proper knowledge. One important IR paradigm is dense retrieval, where the query and candidates, i.e. document, are represented as embeddings, and their semantic relationship can be reflected by the embedding similarity \cite{yates2021pretrained}. With the continual progress on pre-trained model and training algorithm, increasingly powerful embedding models have been developed, such as DPR \cite{dpr}, Contriever \cite{DBLP:journals/tmlr/IzacardCHRBJG22}, GTR \cite{DBLP:conf/emnlp/Ni0LDAMZLHCY22}, E5 \cite{wang2022text}, BGE \cite{bge-xiao2023c}, etc., which substantially improves the quality and universality of dense retrieval. 

Most of the existing dense retrieval models are text-oriented, which can only deal with the data presented in human language. However, a large portion of the world knowledge naturally contains both text and image, e.g., web articles with visual illustration \cite{chang2022webqa}; meanwhile, people's queries can also be flexibly expressed with multiple data modalities, e.g., search queries with exemplar images \cite{cirr-liu2021image,fashioniq-wu2021fashion}. Despite the development of visual-language representation models (VLM), like CLIP \cite{clip-radford2021learning} and ALIGN \cite{jia2021scaling}, the above problem is still challenging in many perspectives. On one hand, the existing VLMs are severely limited in text representation capability, whose retrieval performance is far behind the recent text-only embedding models, like E5 and BGE. On the other hand, the existing VLMs focus more on the independent encoding of text and image; nevertheless, the joint representation of image-text data (e.g., documents with illustrations) is largely unexplored.

In this work, we propose \underline{VIS}ualized \underline{T}ext embedding for universal multi-modal retriev\underline{A}l, namely \textbf{VISTA}. It takes the best of the existing text encoder and image encoder where high-quality multi-modality embedding can be generated from it. In particular, our work presents the following three technical contributions.

First of all, we come up with a flexible {model architecture} to facilitate the generation of multi-modal embedding. It is built upon a powerful and well-trained text encoder, which exhibits proficient text retrieval capability. Meanwhile, it makes the incorporation of visual tokens generated by an expressive image encoder, thereby augmenting the capability of image processing. Such an architecture brings forth two important advantages. 1) It establishes the \textit{in-depth fusion} of text and image data, which substantially contributes to the quality of multi-modal embedding. 2) It also enables the \textit{preservation of the original performance} of text embedding, as the text encoder is fully fixed while the visual tokens are incorporated. 

Secondly, we propose two innovative pipelines for the automatic generation of \textit{Image-Text Composed datasets}, thereby securing large-scale, high-quality data for the training of multi-modal embedding models. These pipelines are designed to cater to scenarios where either the {query} or the {candidate} comprises image-text pairs, thereby facilitating the model to adapt to a diverse range of multi-modal retrieval situations. 

Thirdly, we design a two-stage training algorithm to learn the multi-modal embedding model. Initially, we perform the basic text-to-image matching task with massive weakly-labeled cross-modal data~\cite{laion5b-DBLP:conf/nips/SchuhmannBVGWCC22}, which aligns the visual token embedding with the text encoder. Subsequently, we perform composed text\&image matching with our generated composed image-text datasets, which establishes the multi-modal representation capability for the embedding model.

VISTA is empirically verified by comprehensive experiments. Particularly, it achieves superior performance across various multi-modal retrieval tasks in both zero-shot and supervised settings. Without any task-specific optimization, VISTA is able to outperform or match the leading approach in every downstream evaluation scenario. Besides, VISTA's performance can also be substantially improved if it is continually fine-tuned for corresponding tasks. 

 

\section{Related Work}

\subsection{General Text Embedding}
General text embedding plays an important role in various applications such as web search, question answering~\cite{dpr}, and retrieval augmented generation for large language models~\cite{DBLP:conf/nips/LewisPPPKGKLYR020, DBLP:conf/icml/BorgeaudMHCRM0L22, DBLP:journals/corr/abs-2301-12652}. In recent, numerous effective general text embedding models have been developed, including Contriever~\cite{DBLP:journals/tmlr/IzacardCHRBJG22}, Sentence-Transformer~\cite{DBLP:conf/emnlp/ReimersG19}, OpenAI text embedding~\cite{DBLP:journals/corr/abs-2201-10005}, BGE~\cite{bge-xiao2023c}, and M3~\cite{bge-m3}, etc. 
These models have demonstrated impressive generalizability and robust performance in the realm of text retrieval. However, they exhibit limitations when it comes to handling multi-modal data. This becomes particularly salient with the rising popularity of multi-modal retrieval~\cite{chang2022webqa,CoIR-vo2019composing, remuq-DBLP:conf/acl/0003FGYB23} and multi-modal retrieval-augmented generation~\cite{murag-DBLP:conf/emnlp/ChenHCVC22, racm3-DBLP:conf/icml/YasunagaAS0LLLZ23}.

\subsection{General Multi-Modal Embedding}

Multi-modal retrieval, characterized by queries and/or candidates composed of image-text data, is gaining increasing popularity in practice~\cite{CoIR-vo2019composing, chang2022webqa, remuq-DBLP:conf/acl/0003FGYB23}. Different from cross-modality retrieval models~\cite{clip-radford2021learning} which independently process image and text modalities, multi-modal retrieval necessitates models to have an in-depth understanding of the composed image-text data.
Most existing models for multi-modal embedding primarily rely on the pre-trained CLIP~\cite{clip-radford2021learning} or BLIP~\cite{li2022blip}. For instance, models such as UniVL-DR~\cite{univl-liu2022universal}, Clip4Cir~\cite{clip4cirv3-baldrati2023composed}, and UniIR~\cite{mbeir-wei2023uniir} initially encode image and text separately using the corresponding encoders from CLIP or BLIP. These models then employ a fusion strategy, such as score fusion, to integrate features from both modalities. 

However, these models lack in-depth image-text fusion mechanisms~\cite{mbeir-wei2023uniir, univl-liu2022universal} or are designed for specific tasks~\cite{univl-liu2022universal, saito2023pic2word, clip4cirv3-baldrati2023composed}, rather than for a broad spectrum of multi-modal embedding applications.
Furthermore, the text embedding capabilities of CLIP and BLIP are not on par with recent general text embedding models, which can potentially compromise their performance in tasks that involve processing text-heavy multi-modal documents~\cite{chang2022webqa,remuq-DBLP:conf/acl/0003FGYB23}. A concurrent work, Marvel~\cite{marvel-zhou2023unlock}, leverages pre-trained text embedding models as a foundation for encoding composed image-text documents, facilitated by a visual plugin. However, Marvel is a task-specific model trained for multi-modal document retrieval~\cite{chang2022webqa, univl-liu2022universal}, and it cannot be utilized as a general multi-modal embedding model to handle other tasks, such as composed image retrieval.

\section{VISTA Model}

\subsection{Model Architecture}

\begin{figure}[tb]
    \centering
    \includegraphics[scale=0.7]{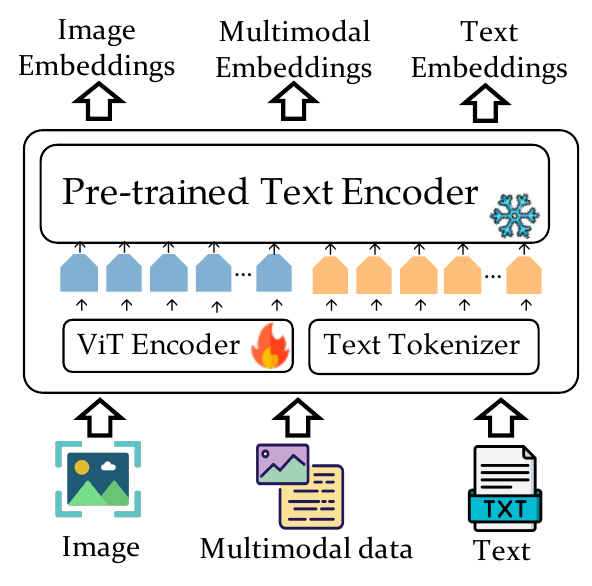}
    \caption{The model architecture of our VISTA model. We use the pre-trained language model as the foundation, making the ViT encoder transfer the Image to recognized tokens of the text encoder.}
    \label{fig:modelarch}
\end{figure}

The core idea of our VISTA is the use of the ViT encoder as an image tokenizer for the text encoder. This enables VISTA to encode a variety of data types, including images, text, and composed image-text data. As shown in~\Cref{fig:modelarch}, we treat the Vision Transformer (ViT)~\cite{vit-DBLP:conf/iclr/DosovitskiyB0WZ21} as an image tokenizer of the text encoder, which allows the pre-trained text model to recognize image tokens while remaining frozen. The benefit of this approach is that it facilitates an \textit{in-depth fusion} of text and image data, while the text encoder retains its robust text embedding capabilities.

Specifically, VISTA encodes text data directly using the pre-trained text encoder, as illustrated by the following formula:
\begin{equation}
\label{eq-text}
\textbf{e}_t = Bert(\{t_0, ..., t_m\})    
\end{equation}
Here, $Bert$ represents the text encoder~\cite{bert-devlin2018bert} and is initialized with a pre-trained general text embedding model. $\{t_0, ..., t_m\}$ and $\textbf{e}_t$ denote the text sequence and its corresponding text embedding, respectively. Notably, we utilize the normalized hidden state of Bert's special token, \texttt{[CLS]}, as the output of the embedding. For image data, the encoding process is defined as follows:
\begin{equation}
\label{eq-image}
\begin{aligned}
&\{\epsilon_0, ..., \epsilon_n\} = ViT(\{i_0, ..., i_n\}) \\
&\textbf{e}_i = Bert(\{\epsilon_0, ..., \epsilon_n\})
\end{aligned}
\end{equation}
where $ViT$ is a vision transformer serving as an image tokenizer, $\{i_0, ..., i_n\}$ is the token sequence of the input image patches, while $\{\epsilon_0, ..., \epsilon_n\}$ corresponds to the sequence of hidden states for image tokens, as produced by $ViT$. The image token sequence $\{\epsilon_0, ..., \epsilon_n\}$ is then encoded by $Bert$ to derive the corresponding image embedding $\textbf{e}_i$. For the composed image-text data, we encode it as:
\begin{equation}
\label{eq-hybrid}
\begin{aligned}
&\{\epsilon_0, ..., \epsilon_n\} = ViT(\{i_0, ..., i_n\}) \\
&\textbf{e}_h = Bert(\{\epsilon_0, ..., \epsilon_n\};~\{t_0, ..., t_m\})
\end{aligned}
\end{equation}
We concatenate the sequence $\{\epsilon_0, ..., \epsilon_n\}$ and $\{t_0, ..., t_m\}$ together, forming an interleaved sequence of image and text tokens. This interleaved sequence is then encoded by $Bert$ to yield the hybrid multi-modal data representation $\textbf{e}_h$.

We exclusively trained $ViT$ during the training procedure while maintaining the text encoder $Bert$ in a frozen state. This strategy is adopted to preserve the powerful text embedding capabilities of the pre-trained text general embedding model.

\begin{figure*}[tb]
    \centering
    \includegraphics[scale=0.5]{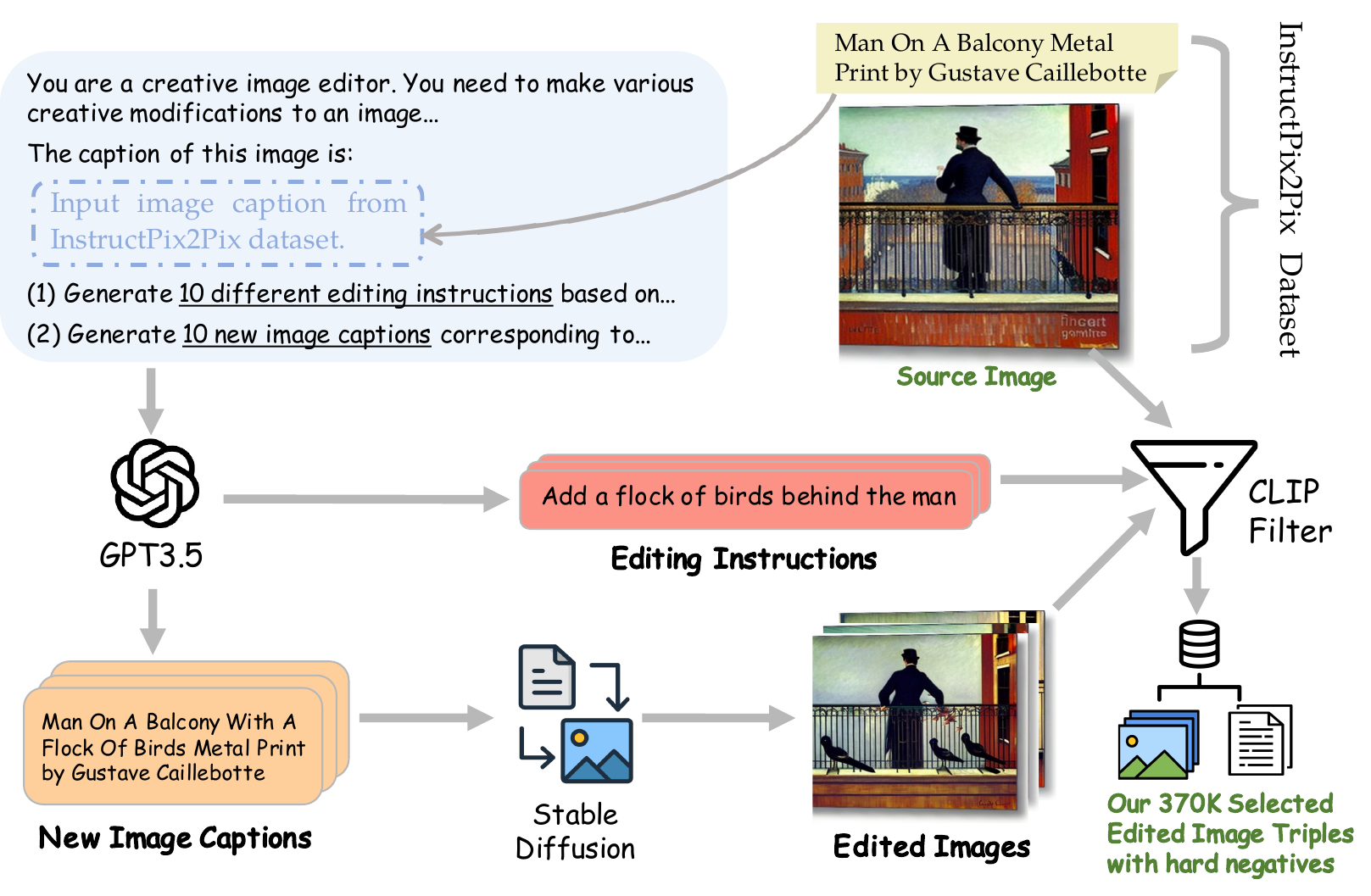}
    \caption{The construction pipeline of Image\&Text To Image (IT2T) dataset.}
    \label{fig:s2datia-edit}
\end{figure*}

\subsection{Data Construction} 
\label{sec:stage2data}

Existing hybrid multi-modal datasets predominantly require human annotation, such as writing queries for multi-modal document retrieval~\cite{chang2022webqa}, annotating semantic relations for composed image retrieval~\cite{cirr-liu2021image, fashioniq-wu2021fashion}, and creating questions and answers for knowledge retrieval~\cite{remuq-DBLP:conf/acl/0003FGYB23}. These costly human annotations limit the scale of hybrid multi-modal datasets, posing challenges for training multi-modal embedding models. To address these challenges, we have designed two pipelines to generate hybrid multi-modal data. These pipelines, based on the scenarios where either the \textit{\textbf{query}} or the \textit{\textbf{candidate}} is composed of image and text, provide a versatile training dataset that can accommodate diverse multi-modal retrieval situations. Our pipelines facilitate the production of two large-scale multi-modal embedding training datasets. 
The statistical information of our generated dataset is presented in~\Cref{tab:gendata}.

\begin{table}
    \centering
    \footnotesize
    \begin{tabular}{lcc}
    \toprule
     Dataset    & Queries & H. Annot. \\
     \midrule
     CIRR~\cite{cirr-liu2021image}          & 36K    & \pmb{$\checkmark$}    \\
     FashionIQ~\cite{fashioniq-wu2021fashion}     & 30K    & \pmb{$\checkmark$}    \\
     \textbf{Our IT2I Data} & 307K   & \pmb{$\times$}         \\
     \midrule
     WebQA~\cite{chang2022webqa}      & ~~21K$^\star$  & \pmb{$\checkmark$} \\
     \textbf{Our T2IT Data}     & 213K   & \pmb{$\times$}            \\
     \bottomrule
    \end{tabular}
    \caption{Comparison of our generated datasets with existing datasets. Queries refers to the number of query-candidate pairs. H. Annot. denotes the necessity of human annotation. $^\star$ The number of queries corresponding to multi-modal documents in WebQA.}
    \label{tab:gendata}
\end{table}

\subsubsection{Image\&Text To Image (IT2I) Dataset}

Inspired by InstructPix2Pix~\cite{brooks2023instructpix2pix}, which devises a synthetic image-editing dataset for image editing models, we establish a pipeline for creating a dataset that is characterized by composed image-text queries. As shown in~\Cref{fig:s2datia-edit}, we feed the caption of the source image $\mathcal{C}_s$ into GPT-3.5~\cite{chatgpt3.5}, prompting it to generate multiple distinct editing instructions $\{\mathcal{T}^1, ..., \mathcal{T}^m\}$ along with their corresponding image captions $\{\mathcal{C}_t^1, ..., \mathcal{C}_t^m\}$, which are then fed into the stable diffusion model~\cite{stablediffusion-rombach2022high} to generate the edited images $\{\mathcal{I}_t^1, ..., \mathcal{I}_t^m\}$. We designate different edited images $\mathcal{I}_{t}^i$ originating from the same source image $\mathcal{I}_s$ as hard negatives for each other. Consequently, we obtain multiple triples $(\mathcal{I}_s, \mathcal{T}^i, \mathcal{I}_{t}^i)$, where $\mathcal{I}_s$ and $\mathcal{T}^i$ are the composed image-text query, and $\mathcal{I}_t^i$ is the target image. We further employ CLIP~\cite{clip-radford2021learning} to filter these triples, resulting in 307K query-candidate pairs with hard negatives. 

A major distinction is that our approach generates multiple editing instructions for each source image, while InstructPix2Pix provides only a single editing instruction per source image. Different edited images can work as hard negatives with each other. Therefore, it prevents the training task from collapsing into a naive image-to-image matching task, which enables the model to jointly understand the image and text data. 


\subsubsection{Text To Image\&Text (T2IT) Dataset} 
\label{sec:datagen-mmdoc} 

We establish another pipeline to construct a pseudo multi-modal document retrieval dataset, in which the candidates are composed of both images and text. Our pipeline operates on a highly descriptive image captioning dataset ShareGPT4V~\cite{sharegpt4v}. ShareGPT4V is characterized by the detailed textual image description that includes multi-granular information, encompassing world knowledge, properties of objects, spatial relationships, etc.

Specifically, for each image $\mathcal{I}$ accompanied by a descriptive caption $\mathcal{C}$, we first input $\mathcal{C}$ into GPT-3.5 and prompt it to generate an article $\mathcal{T}$ that is related to a subtopic of the image. Consequently, we obtain a multi-modal document candidate, denoted as $D=(\mathcal{I}, \mathcal{T})$. We then prompt GPT-3.5 to generate a query ${Q}$ for the generated multi-modal document ${D}$. Through this process, we obtain over 213K triples $({Q}, \mathcal{I}, \mathcal{T})$, where ${Q}$ represents the query and $(\mathcal{I}, \mathcal{T})$ forms the multi-modal document candidate. We demonstrate that the data generated by this simple yet effective pipeline exhibits superior generalization capabilities compared to the manually annotated WebQA~\cite{chang2022webqa} when used to train multi-modal embedding models, as detailed in Section~\ref{sec:abl-data}. More details of the data generation process are shown in~\Cref{appen-datagen}.

\begin{table*}
    \centering
    \begin{tabular}{llccl}
        \toprule
        Dataset & Task & Query Count & Corpus Size & Domain\\
        \midrule
        WebQA~\cite{chang2022webqa} & $q_t\rightarrow c_{t}/c_{it}$ & 4,966   & 944,766 & Wikipedia \\
        CIRR~\cite{CoIR-vo2019composing}& $q_{it}\rightarrow c_i$  &  4,181 & 21,551 &  Open-Domain  \\
        FashionIQ~\cite{fashioniq-wu2021fashion}& $q_{it}\rightarrow c_i$  &  6,016  & 74,381 & Fashion Products \\
        ReMuQ~\cite{remuq-DBLP:conf/acl/0003FGYB23}& $q_{it}\rightarrow c_{t}$ &  3,609 & 195,387 & Wikipedia \\
        OVEN-QS~\cite{OVEN-DBLP:conf/iccv/HuLCKJLTC23}&$q_{it}\rightarrow c_{t}/c_{it}$  & 3,291  & 6,084,491 & Wikipedia \\
        \bottomrule
    \end{tabular}
    \caption{Statistical information for the zero-shot multi-modal retrieval benchmark datasets. $q$ and $c$ represent {query} and {candidate} respectively, with the subscripts $i$, $t$, and ${it}$ denoting image, text, and composed image-text data respectively. During the zero-shot evaluation, we utilize the queries from the validation or test set of each dataset to perform retrieval assessments within the entire corpus of the respective dataset.}
    \label{tab:zs-benchmark}
\end{table*}

\subsection{Two-Stage Training}
\label{sec:trainingstage}
We develop a two-stage training strategy to facilitate the text encoder's ability to encode both image and hybrid multi-modal data into a unified embedding space. We initialize the text encoder with a general embedding model BGE-Base-v1.5~\cite{bge-xiao2023c} and initialize the ViT Encoder with EVA-CLIP-02-Base~\cite{evaclip-sun2023eva}. 

\noindent\textbf{Stage 1: Cross-Modal Training.} In the first training stage, we conduct contrastive language-image pre-training~\cite{clip-radford2021learning} to our VISTA. All training data are uni-modal in this stage, and we utilize the Laion-2B~\cite{laion5b-DBLP:conf/nips/SchuhmannBVGWCC22} for in-depth alignment training, thereby transforming the ViT encoder into a high-quality image tokenizer for the general text embedding model. The training objectives are as follows:
\begin{equation}
\label{eq-loss1-1}
    \min_{\{\theta_I\}}~\mathcal{L}_{s1} = \mathcal{L}_{con}(\textbf{e}_t, \textbf{e}_i) + \mathcal{L}_{con}(\textbf{e}_i, \textbf{e}_t)
\end{equation}
where $\theta_I$ is the parameters of $ViT$. $\mathcal{L}_{con}(\textbf{e}_t, \textbf{e}_i)$ and $\mathcal{L}_{con}(\textbf{e}_i, \textbf{e}_t)$ are bidirectional cross-modal contrastive learning losses, and $\mathcal{L}_{con}(\textbf{u}, \textbf{v})$ can be formulated as:
\begin{equation}
\label{eq-loss1-2}
\mathcal{L}_{con}(\textbf{u}, \textbf{v}) = -\frac{1}{\left |\mathcal{B}\right |} \sum_{i\in \mathcal{B}}^{} \frac{exp(\textbf{u}^{T}_{i}\textbf{v}_{i}/\tau )}{\sum_{j\in \mathcal{B}}^{}exp(\textbf{u}^{T}_{i}\textbf{v}_{j}/\tau ) }    
\end{equation}
where $\mathcal{B}$ represents the set of in-batch samples, and $\tau$ is the temperature parameter that controls the strength of penalties on negative samples. Following the first stage of training, the image tokenizer develops the ability to encode image tokens in a format that the text encoder can interpret.

\noindent\textbf{Stage 2: Multi-Modal Training.} After the first stage of training, the text encoder has gained the ability to independently process image and text modalities, and align them into a unified embedding space. This has laid the groundwork for encoding interleaved sequences of text and image. Building upon this foundation, we further train VISTA to enhance its multi-modal encoding capabilities.

Specifically, we utilize our generated IT2I and T2IT datasets, as constructed in~\Cref{sec:stage2data}, for multi-modal training. The training objective of both the two tasks can be formulated as:
\begin{equation}
\label{eq-loss2}
    \min_{\{\theta_I\}}~\mathcal{L}_{s2} = \mathcal{L}_{con}(\textbf{q}, \textbf{c})
\end{equation}
where $\textbf{q}$ and $\textbf{c}$ represent the embeddings of the query and candidate of these two tasks, respectively. We discover that a 600-step training process on our generated multi-modal training dataset is sufficient to equip VISTA with robust multi-modal embedding capabilities. This not only underscores the effectiveness of our model architecture but also validates the utility of our generated composed image-text training datasets. For more details on training and the hyper-parameter settings used, please refer to~\Cref{appen-impdetail}.

\section{Experimental Results}

\begin{table*}[]
\centering
\begin{tabular}{lccccccc}
\toprule
Models   &\# Params  & WebQA          & CIRR           & FashionIQ     & OVEN-QS       & ReMuQ & Average        \\
\midrule
CLIP       & 149M & 10.54          & 13.37          & 3.56          & 1.06          & 65.05 & 18.72         \\
CLIP-MM    & 149M  & 28.77          & 19.64          & 5.55          & 0.40           & 58.86 & 22.64         \\
\midrule
BLIP      & 224M  & 10.03          & 8.25           & 1.50           & 0.06          & 1.25  & 4.22          \\
BLIP-MM   & 224M  & 30.11          & 10.31          & 1.23          & 0.27          & 55.66 & 19.52         \\
\midrule

Pic2Word  & 224M  & 12.72          & 23.42          & \textbf{8.24}          & 0.97          & 68.99 & 22.87         \\
Pic2Word-MM & 224M   & 24.15          & \textbf{26.09} & \underline{7.65} & 0.82          & 78.09 & 27.36          \\
\midrule
VISTA (Ours)  & 196M     & \textbf{60.11} & 22.51          & \underline{7.51} & \textbf{8.39} & \textbf{84.73} & \textbf{36.65} \\
\bottomrule
\end{tabular}
\caption{Zero-shot evaluation results with Recall@5 on various hybrid multi-modal retrieval benchmarks. The '-MM' notation indicates baseline models that have undergone multi-modal training on our generated data. For zero-shot evaluation, we utilize the entire corpus of each dataset, encompassing all data splits, as the candidate pool.}
\label{tab:zs-main}
\end{table*}

We carry out both zero-shot evaluations and supervised fine-tuning across various benchmarks to substantiate the efficacy and versatility of our VISTA model. In addition, we perform comprehensive ablation studies to scrutinize both the design of the VISTA model and the effectiveness of our stage-2 training datasets.
\subsection{Zero-Shot Retrieval Performance}
\label{sec:exp-zs}

\noindent\textbf{Benchmarks.}
We collect five distinct datasets, encompassing four different multi-modal retrieval tasks. To construct a challenging zero-shot evaluation setup, we perform our evaluation on the entire corpus of each dataset. The overall statistical information is shown in~\Cref{tab:zs-benchmark}, while the detailed information for each benchmark can be found in~\Cref{sec:appen-bench}. 

\noindent\textbf{Metrics.} We uniformly employ Recall@5 as the evaluation metric across all datasets. We employ the dense retrieval approach to evaluate all models across various dataset benchmarks. Each model encodes the query and candidate items into corresponding embedding spaces, and retrieval is performed based on cosine similarity scores using FAISS~\cite{faiss-johnson2019billion}.

\noindent\textbf{Baselines.} We benchmark our VISTA model against three established baseline models: CLIP-B\footnote{https://huggingface.co/openai/clip-vit-base-patch16}~\cite{clip-radford2021learning}, BLIP-B~\cite{li2022blip}, and Pic2Word~\cite{saito2023pic2word}. We utilize the strategies outlined in~\cite{mbeir-wei2023uniir, univl-liu2022universal,saito2023pic2word} to encode composed image-text data for these baseline models. Further details can be found in~\Cref{appn-baseline}. 
Furthermore, to validate the universality of our generated composed image-text dataset, we apply the multi-modal training (\Cref{sec:trainingstage}) to all baseline models. These baseline models $\mathcal{X}$ are denoted as $\mathcal{X}$-MM.

\noindent\textbf{Overall Performance.}
The zero-shot performance of various models is presented in~\Cref{tab:zs-main}. Our VISTA model achieves state-of-the-art average performance across all tasks, with more than 9\% improvement on Recall@5. On the WebQA, OVEN-QS, and ReMuQ datasets, our model significantly outperforms all baselines in zero-shot retrieval performance. While the performance of our model on the composed image retrieval task on CIRR and FashionIQ is slightly lower than the proprietary model, pic2word, it should be noted that pic2word is a model specifically designed for this task. These results affirm the versatility and efficacy of VISTA in hybrid multi-modal retrieval. In addition, through multi-modal training on our generated dataset, we have seen a significant improvement in the zero-shot performance of all baseline models across various tasks. This demonstrates the efficiency and universality of our generated dataset. In addition, the qualitative zero-shot retrieval results of VISTA can be found in~\Cref{sec:appen-zsvis}.

\subsection{Supervised Fine-Tuning Performance}
We fine-tune our VISTA model across a variety of hybrid multi-modal retrieval benchmarks, including WebQA, CIRR, and ReMuQ. During the supervised fine-tuning process, we train all parameters of VISTA. Importantly, we abstain from making any task-specific modifications to VISTA and do not utilize any additional training data. The experimental results demonstrate the robustness and exceptional adaptability of VISTA across various hybrid multi-modal retrieval tasks.

\subsubsection{Fine-Tuning Performance on WebQA}
\textbf{Details \& Metrics.} Following the approach of \cite{univl-liu2022universal}, we fine-tune our VISTA on the training set of WebQA~\cite{chang2022webqa}. We employ hard negatives from \cite{univl-liu2022universal} for training and set the count of hard negatives to 9. During fine-tuning, we set the batch size to 288, and the initial learning rate to 2e-5, and fine-tune for 700 steps. During testing, we use the validation query set to retrieve from the entire corpus. Recall@5/10/20 and MRR@10 serve as our evaluation metrics.

\noindent\textbf{Results.} The experimental results are presented in~\Cref{tab:webqa}. 
VISTA achieves 70.8\% in Recall@5 and 71.0\% in Recall@10. VISTA outperforms the previous SOTA method~\cite{univl-liu2022universal} by over 6\% and exceeded the concurrent work Marvel~\cite{marvel-zhou2023unlock} by more than 4\%.

\begin{table}[]
\centering
\footnotesize
\begin{tabular}{l|cccc}
\toprule
   Methods  & R@5     & R@10    &R@20   & MRR@10 \\
\midrule
CLIP-DPR    & 49.6    & 60.1    & 70.2       & 50.6    \\
UniVL-DR    & 64.5    & 72.9    & 78.8       & 66.8    \\
Marvel-DPR  & 60.1    & 69.6    & 78.0       & 61.6    \\
Marvel-ANCE & 70.8    & 78.8    & 84.3       & 71.0    \\
\midrule
VISTA (Ours) & \textbf{74.9}  & \textbf{83.7}  & \textbf{89.4}  & \textbf{75.3}   \\
\bottomrule
\end{tabular}
\caption{Supervised fine-tuning results on the WebQA dataset. The baseline models CLIP-DPR and UniVL-DR are taken from \cite{univl-liu2022universal}, while Mravel-DPR and Marvel-ANCE are taken from \cite{marvel-zhou2023unlock}. All retrievals are performed on the deduplicated corpus.}
\label{tab:webqa}
\end{table}

\subsubsection{Fine-Tuning Performance on CIRR}
\textbf{Details \& Metrics.} Following the common protocols for composed image retrieval, we evaluate the model performance on the test set of the CIRR~\cite{cirr-liu2021image} dataset. CIRR includes two benchmarks: a standard one where the target search space encompasses the entire test corpus, and a fine-grained subset where the search space is limited to a subgroup of six images similar to the query image. We report Recall@5 (R@5) for the former and Recall@1 (R$_{sub}$@1) for the latter, and calculate the average of these two recall measures. During fine-tuning, we set the initial learning rate to 2e-5 and the batch size to 720, treating the subgroup as hard negatives for training. The model is fine-tuned for a total of 900 steps.

\noindent\textbf{Results.} The experimental results are shown in~\Cref{tab:cirr}. Our VISTA achieves 76.1\% in Recall@5, 75.7\% in R$_{sub}$@1, and an overall average performance of 75.9\% on the test set. Without employing any task-specific module, our VISTA achieves performance on par with state-of-the-art models that have been pre-trained or specifically designed for composed image retrieval.

\begin{table}[]
\centering
\footnotesize
\begin{tabular}{l|ccc}
\toprule
 Methods           & R@5   & R$_{sub}$@1 & Avg. \\
\midrule
CIRPLANT~\cite{cirr-liu2021image}    & 52.6 & 39.2       & 45.9    \\
CompoDiff~\cite{gu2023compodiff}   & 54.4 & 35.8      & 45.1    \\
Combiner~\cite{combiner-baldrati2022effective}    & 65.4 & 62.4      & 63.9   \\
Blip4CIR+Bi~\cite{blip4cirliu2024bi} & 73.1 & 72.1       & 72.6   \\
CLIP4Cir~\cite{clip4cirv3-baldrati2023composed} & \underline{77.0} & 73.2      & 75.1    \\
CoVR~\cite{ventura2023covr}        & \textbf{78.6}  & 75.0      & \textbf{76.8}   \\
\midrule
VISTA (Ours)        & 76.1  & \textbf{75.7}     & \underline{75.9}  \\
\bottomrule
\end{tabular}
\caption{Supervised fine-tuning results on the CIRR test set.}
\label{tab:cirr}
\end{table}

\begin{table}[]
\centering
\begin{tabular}{l|cc}
\toprule
Methods   & R@5   & R@10  \\
\midrule
BM25~\cite{bm25-robertson2009probabilistic}           &  8.8      & 10.8      \\
DPR~\cite{dpr}       & 43.4  & 48.8  \\
SEAL~\cite{SEAL-bevilacqua2022autoregressive}      & 66.4  & 74.1  \\
ReViz~\cite{remuq-DBLP:conf/acl/0003FGYB23}     & 62.4  & 71.6  \\
ReViz-ICT~\cite{remuq-DBLP:conf/acl/0003FGYB23} & 76.2  & 83.3  \\
GeMKR~\cite{long2024generative}     & 90.3  & 92.7  \\
\midrule
VISTA (Ours)      & \textbf{96.3} & \textbf{97.3} \\
\bottomrule
\end{tabular}
\caption{Supervised fine-tuning results on the ReMuQ test set.}
\label{tab:remuq}
\end{table}

\subsubsection{Fine-Tuning Performance on ReMuQ}
\textbf{Details \& Metrics.} We fine-tune our model on the training set of ReMuQ and test it on its test set. The evaluation is conducted on the entire knowledge base of ReMuQ. We report Recall@5 and Recall@10, and compare our results with state-of-the-art methods. The initial learning rate is set to 2e-5, the batch size to 1,920, and the model is fine-tuned for 200 steps.

\noindent\textbf{Results.} As shown in~\Cref{tab:remuq}, VISTA achieves as high as 96.3\% in Recall@5, surpassing the latest state-of-the-art method~\cite{long2024generative} by more than 5\%. These results demonstrate the powerful capability of VISTA in multi-modal knowledge base retrieval and highlight its considerable potential for application in multi-modal retrieval augmented generation for LLMs.

\begin{table*}[]
\centering
\begin{tabular}{lcccccc}
\toprule
Model       & WebQA          & CIRR           & FashionIQ     & OVEN-QS       & ReMuQ          & Avg.  \\
\midrule
VISTA$_{S1}$            & 35.70           & 9.59           & 1.33          & 3.82          & 21.53          & 14.39 \\
\midrule
w/ InstructPix2Pix      & 44.24          &   14.47             &    2.88           & 5.14     &  80.60        & 29.47 \\
w/ Ours-IT2I       & 51.87          & \underline{21.29}   & \underline{6.73}   & 3.40      & {\underline{\textbf{89.06}}} & 34.47 \\
\midrule
w/ WebQA       &    -           &  10.64              &    2.03           &     3.92          & 77.42         &    -            \\
w/ Ours-T2IT        & \underline{57.28}          & 15.86          & 3.81          & \underline{5.38}          & {74.67}          & 31.40  \\
\midrule
\textbf{VISTA}  & \textbf{60.11} & \textbf{22.51} & \textbf{7.51} & \textbf{8.39} & 84.73          & \textbf{36.65} \\
\midrule
VISTA-SF           & 59.46          & 15.93          & 5.27          & 1.19          & 83.04          & 32.98 \\

\bottomrule
\end{tabular}
\caption{Ablation studies: The zero-shot performance of models that use different stage-2 training data or multi-modal fusion methods. IT2I and T2IT respectively represent our generated Image\&Text To Image Dataset and Text To Image\&Text Dataset. \underline{Underlined values} indicate where the associated training dataset has significantly improved performance on the corresponding benchmarks. VISTA-SF is an ablation model that employs the score-fusion method to encode composed image-text data.}
\label{tab:abl-data}
\end{table*}

\subsection{Ablation Analysis}
\subsubsection{The Impact of Stage-2 Training Data}
\label{sec:abl-data}

\Cref{tab:abl-data} investigates the effect of our generated composed image-text training data (stage-2 training data) on the zero-shot performance across a variety of tasks. VISTA$_{S1}$ denotes the model post the first stage of image-text cross-modal alignment. Building on this model, we use different training data configurations to analyze the influence of our generated stage-2 training data on the multi-modal embedding capabilities of VISTA.

Compared to the VISTA$_{S1}$ model, the use of our generated Image\&Text To Image (IT2I) Dataset leads to significant performance enhancements on the CIRR, FashionIQ, and ReMuQ benchmarks. These benchmarks are characterized by their inherent need for a comprehensive understanding of multi-modal queries. In comparison to training on the InstructPix2Pix dataset~\cite{instructblip-DBLP:journals/corr/abs-2305-06500} that lacks hard negatives, the model trained on our IT2I dataset demonstrates superior multi-modal retrieval performance. The hard negatives that we generate can foster the model's comprehension of the interplay between images and text, thereby preventing the model from excessively relying on image feature similarity rather than semantic correlation during the training process.

In contrast to the VISTA$_{S1}$ model, the employment of our generated Text To Image\&Text (T2IT) dataset notably enhances the zero-shot performance on the WebQA and OVEN-QS benchmarks. These benchmarks require a composed understanding of multi-modal candidates, a demand directly met by the training scenarios presented in our T2IT dataset. Additionally, we evaluate the performance of the model using WebQA for second-stage training. Except for ReMuQ, which is sourced from WebQA, the model exhibits superior performance when trained with our T2IT dataset compared to when it is trained with WebQA. This suggests that our T2IT data offers enhanced generalization capabilities compared to manually annotated datasets when used to train multi-modal embedding models.

In the final Stage-2 training, we conduct dual-task training using both our IT2I and T2IT datasets on the basis of VISTA$_{S1}$, resulting in the development of VISTA. Among all different training data configurations, VISTA achieves the best performance in four out of the five benchmarks. This result confirms the synergistic effect of integrating both datasets in the training of multi-modal embedding models.

\subsubsection{Multi-Modal Fusion Methods}
\Cref{tab:abl-data} also examines the benefits of the VISTA model architecture in encoding composed image-text data in comparison to VISTA-SF (last line). For VISTA-SF, a score-fusion approach is employed during the encoding of multi-modal data. This involves independently encoding images and text using the VISTA model, followed by an element-wise addition of the embeddings derived from both modalities. 
The experimental results demonstrate that VISTA achieves substantial improvement, significantly outperforming the VISTA-SF model. This can be attributed to the fact that score-fusion cannot deeply comprehend the integration of images and text, whereas VISTA is proficient in consistently encoding and interpreting composed image-text data. These findings underscore the advantages of the VISTA model in encoding interleaved text and visual sequences.

\section{Conclusion}
\label{sec:conlusion}
In this paper, we introduce \textbf{VISTA}, an \underline{VIS}ualized \underline{T}ext embedding approach for universal multi-modal retriev\underline{A}l. Our work makes three significant contributions. Firstly, we design a flexible model architecture that enables the in-depth fusion of text and image data, while maintaining the powerful performance of the general text embedding models. Secondly, we develop two data generation strategies for training multi-modal embedding models without the need for manual annotation. Lastly, we introduce a two-stage training algorithm that rapidly enhances the multi-modal representation capability of VISTA. Extensive experimental results demonstrate the superior performance of VISTA in both zero-shot and supervised fine-tuning settings for various multi-modal retrieval tasks.

\section*{Limitations}

As we reflect on the work conducted, we identify two areas of potential refinement in our approach. The first area concerns the diversity of image styles in our Image\&Text To Image (IT2T) dataset. The images are generated using a stable diffusion model, which might have limited the range of styles in the dataset. The second area relates to the handling of image tokens. In our current approach, we feed all image tokens directly into the text encoder. This procedure might inadvertently heighten the computational load due to the uniform sequence length. A potential improvement could be the implementation of variable-length image token sequences, which could reduce sequence lengths and consequently lead to more efficient computation.

\section*{Ethics Statement}
All training data used in our model have undergone rigorous screening to remove harmful content. This includes both the LAION-5B dataset constructed by~\cite{laion5b-DBLP:conf/nips/SchuhmannBVGWCC22} and the multi-modal dataset we built ourselves. Despite our best efforts, we acknowledge that we cannot fully guarantee that these screenings were entirely comprehensive or without omissions. Furthermore, we strongly discourage the use of VISTA for encoding and retrieving sensitive content.

\section*{Acknowledgements}
This research is supported by National Science and Technology Major Project (2023ZD0121504) and National Natural Science Foundation of China (NSFC-62306046, NSFC-62272054).

\bibliography{custom}

\appendix
\section*{Appendix}

\section{More Details of Data Construction}
\label{appen-datagen}
\begin{figure}
    \centering
    \includegraphics[width=1\linewidth]{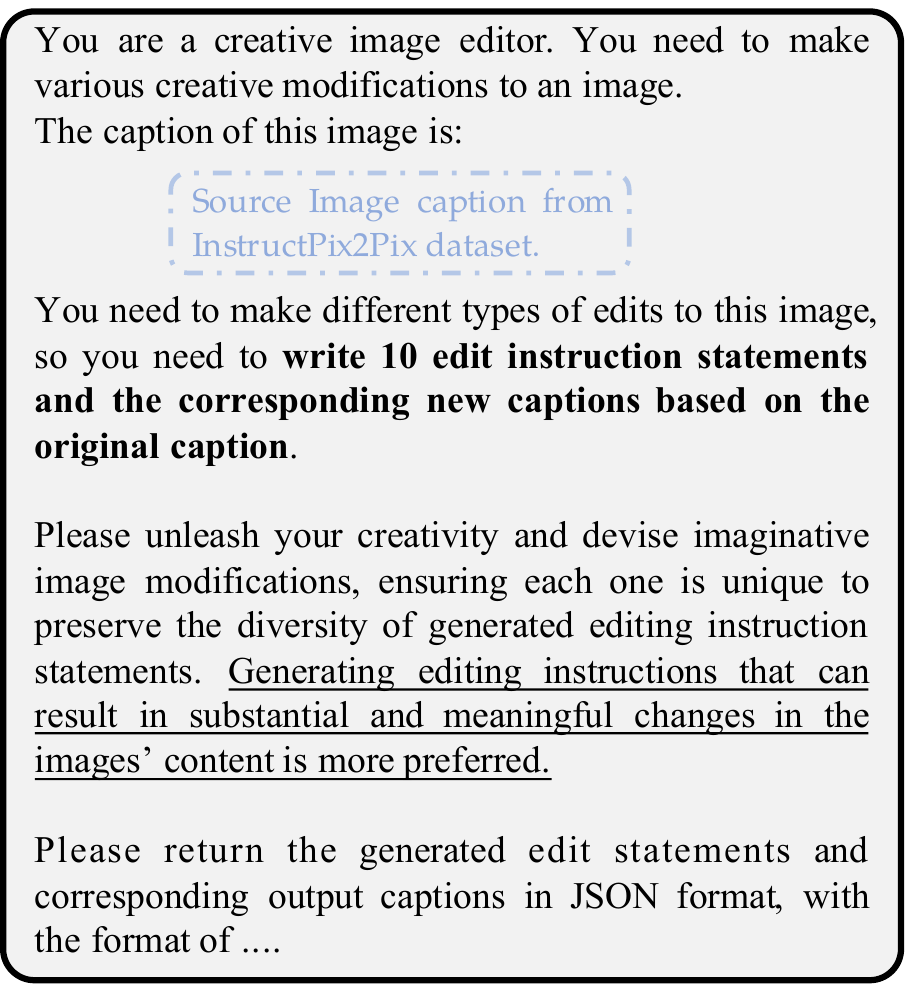}
    \caption{The specific prompts utilized during the generation of the Image\&Text To Image (IT2T) dataset.}
    \label{fig:edit-image-prompt}
\end{figure}

\noindent\textbf{Image\&Text To Image (IT2T) Dataset.} The specific prompts utilized in creating the I2IT dataset are illustrated in~\Cref{fig:edit-image-prompt}. We direct GPT-3.5 to generate a range of image editing instructions that significantly alter the semantic content of the images. This strategy diverges from how image editing datasets for image editing models are constructed. Datasets such as InstructPix2Pix~\cite{brooks2023instructpix2pix} prioritize dealing with the challenges of intricate image edits, whereas our focus with embedding models is more on understanding the relationships among image semantics.

\begin{figure*}
    \centering
    \includegraphics[width=1\linewidth]{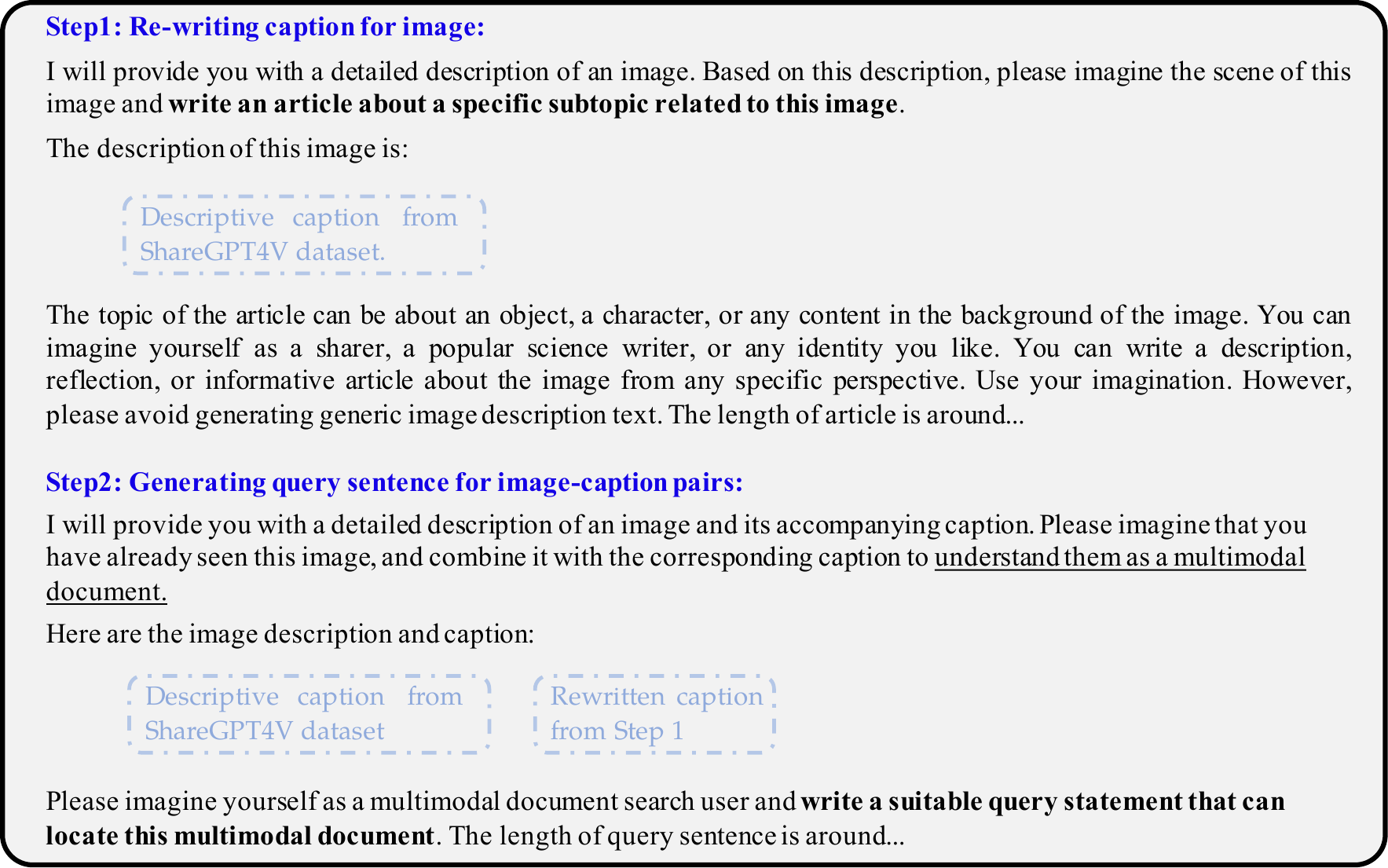}
    \caption{The specific prompts employed in the generation of the Text to Image\&Text (T2IT) dataset, with the lengths of the articles and queries randomly assigned in each data generation iteration to ensure diversity. Typically, articles are approximately 50 words, and queries are within 20 words.}
    \label{fig:mmdoc-prompt}
\end{figure*}

\noindent\textbf{Text to Image\&Text (T2IT) Dataset.} The steps and prompts used in the construction of the T2IT dataset are illustrated in~\Cref{fig:mmdoc-prompt}. As described in~\Cref{sec:datagen-mmdoc}, the generation process of the T2IT dataset is divided into two parts: the first step involves generating an article about the image subtopic, and the second step involves generating a query for the multimodal document. It is worth noting that, in the second step of generating a query for the multimodal document, we still use descriptive captions from ShareGPT4V~\cite{sharegpt4v} to represent the images in the multimodal documents, as GPT-3.5 cannot directly process image data.

\section{More Training Details of VISTA}
\label{appen-impdetail}
In the first training stage, we utilize the FLIP~\cite{flip-li2023scaling} strategy to improve the time efficiency of image-text contrastive training. We randomly mask 50\% of image tokens. This phase is trained for 116K steps. Subsequently, we conduct unmasked tuning~\cite{flip-li2023scaling} for an extra 48K steps. The batch size is 16K throughout this stage. In the second training stage, the quantity of hard negative examples in the IT2I dataset is set to 3. This stage is trained for only 600 steps with a batch size of 1920. Across both stages, we set the temperature coefficient $\tau$ for contrastive learning at 0.02. We initiate the learning rate at $2e-5$ and apply a linear decay strategy for subsequent adjustments. 

\section{Detailed Information of Benchmarks}
\label{sec:appen-bench}
To evaluate the effectiveness of our VISTA model in hybrid multi-modal retrieval tasks, we collected five distinct datasets, encompassing four different multi-modal retrieval tasks. Each dataset features hybrid data in either the query or candidate components, requiring a joint embedding of image and text data. Unless otherwise stated, for each dataset, all zero-shot evaluations are conducted using the dev split as the query set, with retrieval carried out across the entirety of its corpus. 

\noindent\textbf{WebQA}~\cite{chang2022webqa} is a multi-hop and multi-modal open-domain question answering dataset. The \textit{Multi-modal Documents Retrieval} task on WebQA is proposed by~\cite{univl-liu2022universal}. This task involves identifying suitable text or image-text pair candidates based on a query text. Notably, we de-duplicate the corpus of WebQA, following the same procedure as UniIR~\cite{mbeir-wei2023uniir}. The processed corpus encompasses 544,489 unique text documents, and 403,277 distinct image-caption pairs.

\noindent\textbf{CIRR}~\cite{cirr-liu2021image} is an open-domain dataset designed for the \textit{Composed Image Retrieval (CIR)} task~\cite{CoIR-vo2019composing}. In this task, each query comprises an image-text pair, which includes a reference image and instructive text that delineates the differences between the reference and target images. 

\noindent\textbf{FashionIQ}~\cite{fashioniq-wu2021fashion} serves as another relevant dataset for the CIR task, specifically focusing on fashion products, including dresses, shirts, and top-tees.

\noindent\textbf{ReMuQ}~\cite{remuq-DBLP:conf/acl/0003FGYB23} is a dataset curated for \textit{Knowledge Retrieval with Multi-modal Queries} task. In the ReMuQ dataset, each query is composed of an image and an associated textual question, the objective being to search pertinent content from a textual knowledge base. Due to the absence of a dev split, we evaluate ReMuQ on the test split. 

\noindent\textbf{OVEN-QS} is the \texttt{Query Split} of the OVEN benchmark~\cite{OVEN-DBLP:conf/iccv/HuLCKJLTC23}. OVEN-QS is proposed for the \textit{Entity Retrieval with Visually-Situated Queries} task. Each query in OVEN-QS is an image-text pair, with the text component being visually situated and filtered out from VQA scenarios. This dataset necessitates identifying the correct entity from a diverse set of text or image-text candidates.

\section{Implementation details of Zero-Shot Baselines}
\label{appn-baseline}
For both CLIP and BLIP models, we adopt a score fusion approach on image-text pair data as outlined in~\cite{mbeir-wei2023uniir, univl-liu2022universal}. The score fusion process is represented as follows:
\begin{equation}
\label{eq-score-fusion}
\textbf{e}_h = \Phi_T(T) + \Phi_I(I)
\end{equation}
where $\textbf{e}_h$ denotes the embedding of composed image-text data, $T$ and $I$ represent the text and image data respectively, while $\Phi_T$ and $\Phi_I$ refer to the text and image encoders of the CLIP/BLIP models, correspondingly.

For Pic2Word model, a proprietary model dedicated to zero-shot composed image retrieval. We employ this model by mapping an image to a pseudo language token in the text encoder, adhering to their methodology for encoding composed image-text queries:
\begin{equation}
\label{eq-p2w}
\begin{aligned}
&\texttt{[*]} = Img2Txt(\Phi_I(I))    \\
&\textbf{e}_h = \Phi_T(\texttt{"a photo of [*]"}; T) 
\end{aligned}
\end{equation}
In this equation, $Img2Txt$ is the mapping network that translates the image feature into a pseudo language token, denoted as $[*]$. The image and text encoders, $\Phi_I$ and $\Phi_T$ are derived from the CLIP-L model~\cite{clip-radford2021learning}. It derives the composed image-text features by integrating the pseudo language token with the text, subsequently processing this unified input through the text encoder.

\begin{table*}[ht]
\centering
\caption{Influence of token order on VISTA performance in processing interleaved image-text data.}
\label{tab:token-order-results}
\begin{tabular}{lcccccc}
\hline
Token Order & WebQA & CIRR & FashionIQ & OVEN-QS & ReMuQ & Avg. \\
\hline
(visual tokens, text tokens) & 60.11 & 22.51 & 7.51 & 8.39 & 84.73 & 36.65 \\
(text tokens, visual tokens) & 59.49 & 22.10 & 7.36 & 9.54 & 84.68 & 36.63 \\
\hline
\end{tabular}
\end{table*}

\section{The Impact of Token Order}
\label{sec:appen-tokenorder}
As outlined in \Cref{eq-hybrid}, we exclusively utilized the (image tokens, text tokens) order to process the interleaved image-text data in all previous experiments. To investigate the potential impact of token order on the performance of our VISTA, we conducted additional experiments using the (text tokens, image tokens) order. The evaluation results in zero-shot settings are presented in \Cref{tab:token-order-results}, with all reported results based on the Recall@5 metric. The experimental findings indicate that altering the order of image and text tokens does not significantly impact the model's performance.

\section{Zero-Shot Qualitative Results}
\label{sec:appen-zsvis}
\begin{figure*}
    \centering
    \includegraphics[width=1\linewidth]{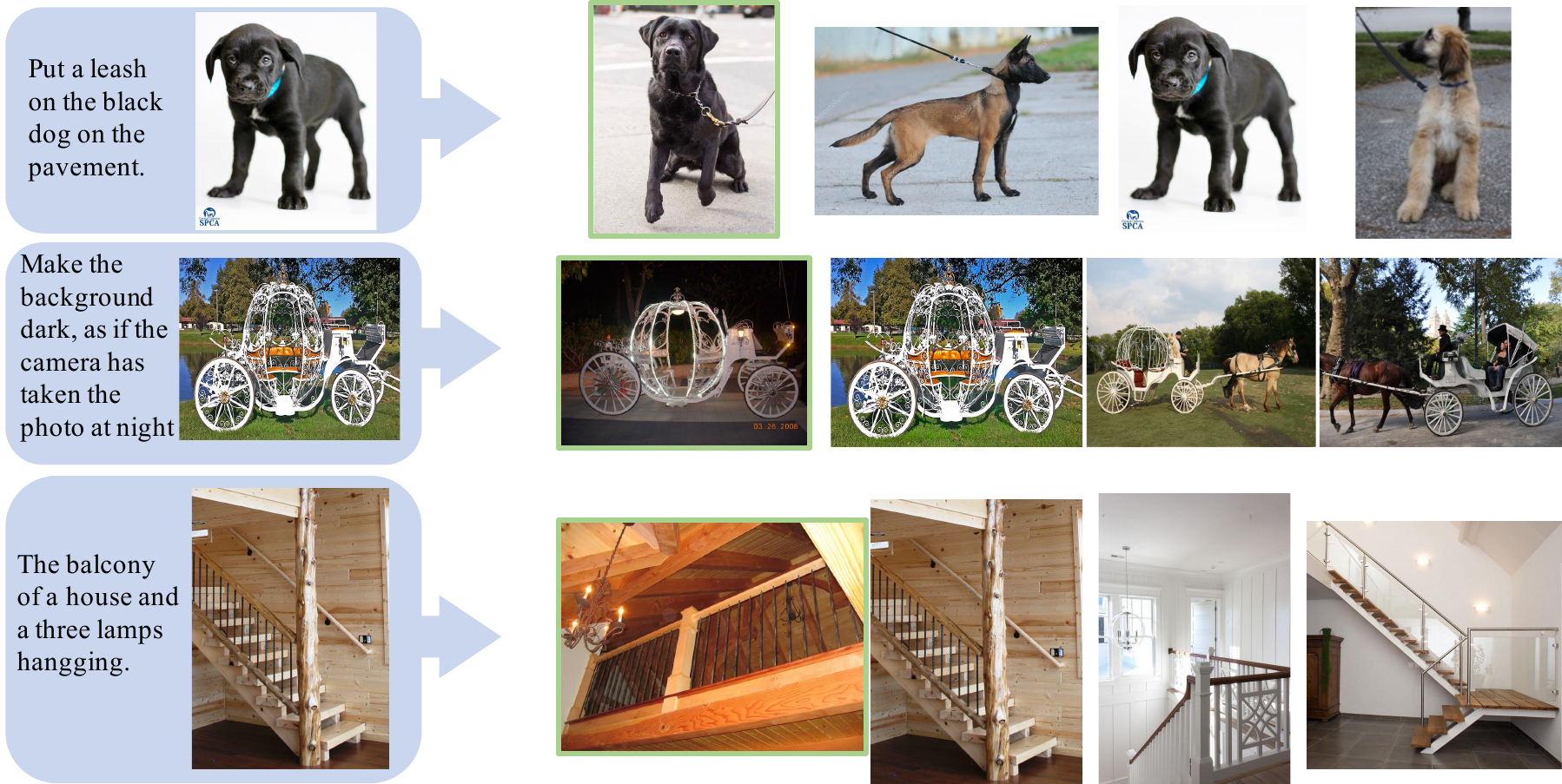}
    \caption{The Qualitative Examples of our VISTA Model on the CIRR Benchmark.}
    \label{fig:vis-cirr}
\end{figure*}

\begin{figure*}
    \centering
    \includegraphics[width=1\linewidth]{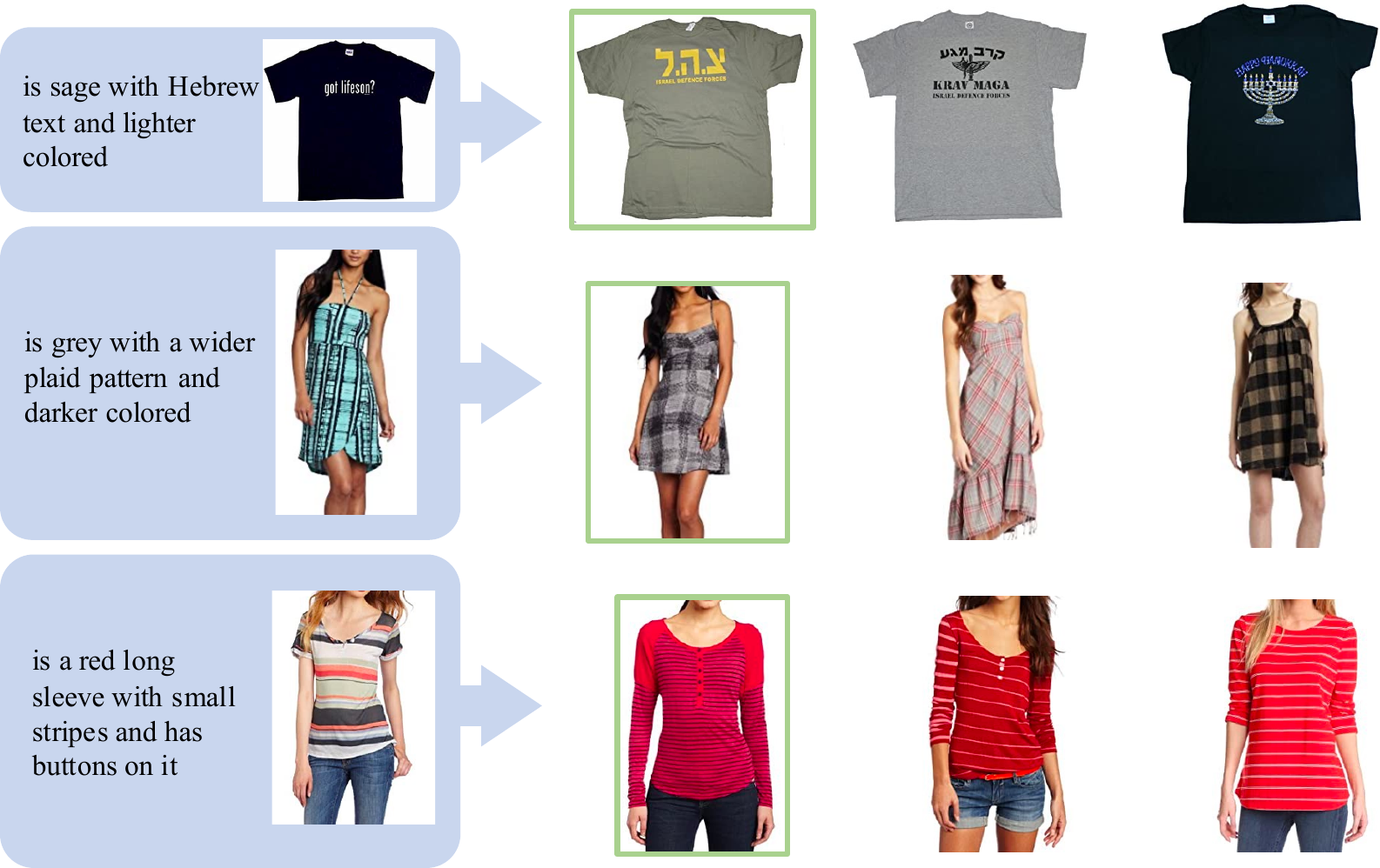}
    \caption{The Qualitative Examples of our VISTA Model on the FashionIQ Benchmark.}
    \label{fig:vis-fashioniq}
\end{figure*}

\begin{figure*}
    \centering
    \includegraphics[width=1\linewidth]{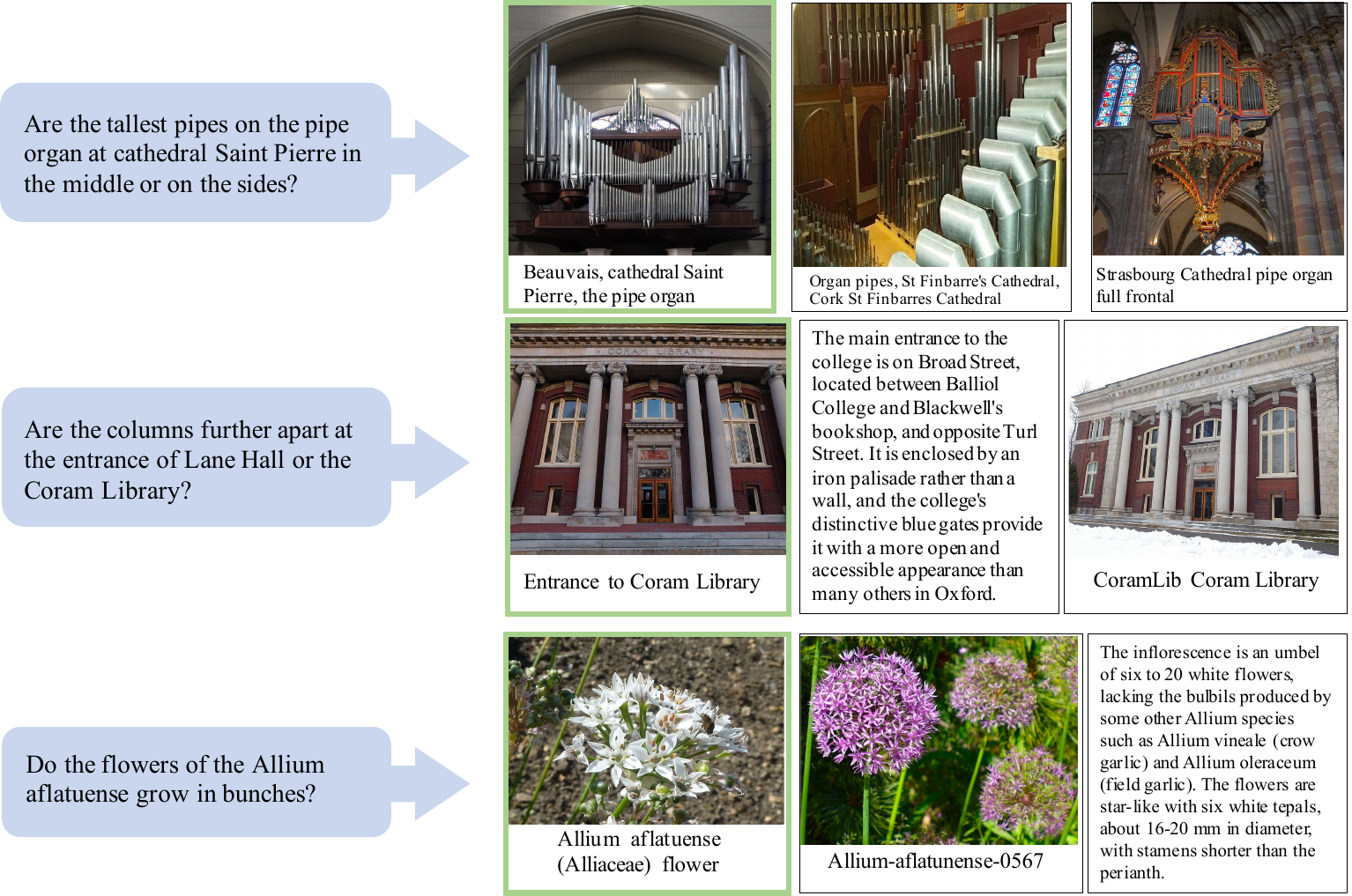}
    \caption{The Qualitative Examples of our VISTA Model on the WebQA Benchmark.}
    \label{fig:vis-webqa}
\end{figure*}

Figures~\ref{fig:vis-cirr}, \ref{fig:vis-fashioniq}, and \ref{fig:vis-webqa} illustrate qualitative zero-shot examples from our VISTA model on the CIRR~\cite{cirr-liu2021image}, Fashion IQ~\cite{fashioniq-wu2021fashion}, and WebQA~\cite{chang2022webqa} datasets, respectively. In each figure, the query is presented on the extreme left, with the top@K retrieval results displayed to its right. The ground-truth candidate, as identified by the benchmark, is denoted by a green box. We conduct our retrieval process across the entire corpus for each dataset, which leads to identifying candidates that meet the retrieval requirements but are not tagged as ground truth. These results demonstrate the impressive zero-shot multi-modal retrieval performance of our VISTA model. 

\end{document}